\documentclass[12pt]{article}
\title{
\vspace{-3mm}
\rightline{\small IFUP-TH 2002/7}
\vspace{8mm}
\bf Deconfining phase transition in the 3D Georgi-Glashow model
with finite Higgs-boson mass}
\author{Dmitri Antonov \thanks{
E-mail: {\tt antonov@df.unipi.it}} \thanks{
Permanent address: ITEP, B. Cheremushkinskaya 25, RU-117 218 Moscow, Russia.}
\\
{\it INFN-Sezione di Pisa, Universit\'a degli studi di Pisa,}\\
{\it Dipartimento di Fisica, Via Buonarroti, 2 - Ed. B -
I-56127 Pisa, Italy}}
\date{}
\begin{document}
\maketitle
\vspace{1mm}
\centerline{\bf {Abstract}}
\vspace{3mm}
\noindent
The (2+1)D Georgi-Glashow model is explored at finite
temperature in the regime when the Higgs boson is not infinitely heavy.
The resulting Higgs-mediated interaction of monopoles leads to the appearance
of a certain upper bound for the parameter of the weak-coupling approximation.
Namely, when this bound is exceeded, the cumulant expansion used for the average over the Higgs
field breaks down.
The finite-temperature deconfining phase transition with the account
for the same Higgs-mediated interaction of monopoles is further analysed. It is demonstrated that
in the general case, accounting for this interaction leads to the existence of two
distinct phase transitions separated by the temperature region where W-bosons exist
in both, molecular and plasma, phases.
The dependence of possible ranges of the critical temperatures corresponding to these
phase transitions on the parameters of the Georgi-Glashow model is discussed.
The difference in the RG behaviour of the fugacity of W-bosons from the respective behaviour
of this quantity in the compact-QED limit of the model is finally pointed out.

\vspace{3mm}

\vspace{10mm}

\section{Introduction}

Although the confining properties of the (2+1)D Georgi-Glashow model are known since
the second half of the seventies~\cite{1}, its finite-temperature properties were addressed
only recently, in refs.~\cite{2}-\cite{nd}.
In ref.~\cite{2}, it has been shown that at the temperature equal to
$g^2/2\pi$, where $g$ stands for the electric coupling constant,
the monopole plasma undergoes the Berezinsky-Kosterlitz-Thouless (BKT)
phase transition~\cite{BKT} to the molecular phase.
In refs.~\cite{W},~\cite{W1}, the relevance of the charged W-plasma to the dynamics
of the deconfining phase transition in the Georgi-Glashow model has been pointed out.
In particular, in ref.~\cite{W}, it has been shown that the phase transition is associated with the
deconfinement of W-bosons, belongs to the
2D-Ising universality class and occurs at the temperature approximately equal to $g^2/4\pi$.
Further, in ref.~\cite{W1}, various physical aspects of this phase transition,
as well as of the analogous transition in the $SU(N)$-generalization of the Georgi-Glashow model
with $N>2$ have been studied. In ref.~\cite{q}, the monopole BKT
phase transition has been explored in the presence of dynamical massless
fundamental quarks. In this way, it has been shown that the presence
of one quark flavour makes the critical temperature of this phase transition equal to $g^2/4\pi$, while
in the presence of more than one flavour, the respective critical temperature becomes exponentially small.

In ref.~\cite{nd}, the influence of the Higgs field to the dynamics of the Georgi-Glashow
model has been studied both at zero and nonzero temperature.
In particular, it has been found that the
finiteness of the Higgs-boson mass does not change the value of the critical temperature
of the monopole BKT phase transition. However, in ref.~\cite{nd} the effects of W-bosons have been disregarded,
that makes the performed analysis incomplete.
The aim of the present letter is to explore the influence of the Higgs-mediated interaction of
monopoles to the deconfining phase transition with the account for W-bosons.
The phase transition occurs when the density of monopoles becomes equal to the one of W-bosons. Up to inessential
subleading corrections, this takes place when the exponent of the monopole
fugacity is equal to that of the fugacity of W-bosons~\cite{W}. [Another way to understand why the phase transition
occurs when the two fugacities are equal to each other is to notice that once this happens,
the thickness of the string confining two W's (which is proportional to
$({\rm monopole}~ {\rm fugacity})^{-1/2}$)
becomes equal to the average distance between the W's (proportional to
(fugacity of W's$)^{-1/2}$).
This qualitative result was also confirmed by the RG analysis performed in ref.~\cite{W}.]
On the other hand, the average over the Higgs field
in the dimensionally-reduced theory (one works with at finite temperatures) changes the monopole fugacity.
Owing to this effect, the critical temperature
of the deconfining phase transition changes as well. Moreover, we shall see that the average over the Higgs field makes
the monopole fugacity temperature-dependent. Due to that, comparison of the exponents of two fugacities yields no more
a single value of the critical temperature, but rather a quadratic equation for this temperature. Consequently, in general,
one gets two distinct critical temperatures. We shall discuss possible
ranges of these temperatures and also the modification
of the RG behaviour of the model due to the existence of two phase transitions instead of one.
Besides that, we shall see that the requirement of convergence
of the cumulant expansion, one should demand in the course of the average over the Higgs field, leads to a certain
upper bound for the parameter of the weak-coupling approximation.

The organization of the letter is the following. In the next Section, we shall consider the dual theory
describing the (2+1)D Georgi-Glashow model at finite temperature and the peculiarities of the average
over the Higgs field in that theory. In Section~3, there will be discussed the deconfining phase transition
and the RG properties of the model.
The main results of the letter will then be summarized
in the Conclusions. In the Appendix~A, some technical details necessary for the evaluation of a certain integral
will be outlined.

\section{3D Georgi-Glashow model at finite temperature beyond the compact-QED limit}

The Euclidean action of the (2+1)D Georgi-Glashow
model reads~\cite{1}

\begin{equation}
\label{GG}
S=\int d^3x\left[\frac{1}{4g^2}\left(F_{\mu\nu}^a\right)^2+
\frac12\left(D_\mu\Phi^a\right)^2+\frac{\lambda}{4}\left(
\left(\Phi^a\right)^2-\eta^2\right)^2\right].
\end{equation}
Here, the Higgs field $\Phi^a$ transforms by the adjoint representation,
$D_\mu\Phi^a\equiv\partial_\mu\Phi^a+\varepsilon^{abc}A_\mu^b
\Phi^c$. Next, $\lambda$ is the Higgs coupling constant of dimensionality [mass],
$\eta$ is the Higgs v.e.v. of dimensionality $[{\rm mass}]^{1/2}$, and
$g$ is the electric coupling constant of the same dimensionality.

At the one-loop level, the sector of the theory~(\ref{GG}) containing
dual photons and Higgs bosons is represented by the following partition function~\cite{dietz}:

\begin{equation}
\label{pf}
{\cal Z}=1+\sum\limits_{N=1}^{\infty}\frac{\zeta^N}{N!}\left(\prod\limits_{i=1}^{N}\int d^3z_i
\sum\limits_{q_i=\pm 1}^{}\right)\exp\left[-\frac{g_m^2}{8\pi}
\sum\limits_{{a,b=1\atop a\ne b}}^{N}\left(\frac{q_aq_b}{|{\bf z}_a-{\bf z}_b|}-
\frac{{\rm e}^{-m_H|{\bf z}_a-{\bf z}_b|}}{|{\bf z}_a-{\bf z}_b|}\right)\right]\equiv
\int {\cal D}\chi{\cal D}\psi {\rm e}^{-S},
\end{equation}
where

\begin{equation}
\label{1}
S=\int d^3x\left[\frac12(\nabla\chi)^2+\frac12(\nabla\psi)^2
+\frac{m_H^2}{2}\psi^2-2\zeta{\rm e}^{g_m\psi}\cos(g_m\chi)\right]
\equiv\int d^3x{\cal L}[\chi,\psi|g_m,\zeta].
\end{equation}
Clearly, this partition function describes the grand canonical ensemble of monopoles with the
account for their Higgs-mediated interaction.
In eqs.~(\ref{pf}) and~(\ref{1}), $\chi$ is the dual-photon field, and the field $\psi$ accounts for the Higgs field,
whose mass reads $m_H=\eta\sqrt{2\lambda}$. Note that from eq.~(\ref{pf}) it is straightforward to deduce that when $m_H$
formally tends to infinity, one arrives at the conventional sine-Gordon theory of the
dual-photon field~\cite{1} describing the compact-QED limit of the model.
Next, in the above equations, $g_m$ stands for the magnetic coupling constant
related to the electric one as $g_mg=4\pi$, and
the monopole fugacity $\zeta$ has the form:
$\zeta=\frac{m_W^{7/2}}{g}\delta\left(\frac{\lambda}{g^2}\right)
{\rm e}^{-4\pi m_W\epsilon/g^2}$.
In this formula, $m_W=g\eta$ is the W-boson mass,
and $\epsilon=\epsilon(\lambda/g^2)$ is a certain monotonic, slowly
varying function, $\epsilon\ge 1$, $\epsilon(0)=1$~\cite{bps},
$\epsilon(\infty)\simeq 1.787$~\cite{kirk}.
As far as the function $\delta$ is concerned,
it is determined by the loop corrections.
It is known~\cite{ks} that this function grows
in the vicinity of the origin [i.e., in the Bogomolny-Prasad-Sommerfield (BPS) limit~\cite{bps}]. However,
the speed of this growth is so that it does not spoil the exponential smallness
of $\zeta$ in the standard weak-coupling regime $g^2\ll m_W$ which
we shall imply throughout this letter.

At finite temperature $T\equiv1/\beta$, one should supply the fields $\chi$ and $\psi$
with the periodic boundary conditions in the temporal direction, with the period equal to
$\beta$. Owing to that, the lines of the magnetic field emitted by a monopole cannot cross
the boundary of the one-period region and consequently, at the distances larger than $\beta$,
should go almost parallel to this boundary, approaching it.
Therefore, monopoles separated by such distances
interact via the 2D Coulomb potential, rather than the 3D one. Since the average distance
between monopoles in the plasma is of the order of $\zeta^{-1/3}$, we see that at $T\ge\zeta^{1/3}$,
the monopole ensemble becomes two-dimensional. Owing to the fact that $\zeta$ is exponentially
small in the weak-coupling regime under discussion, the idea of dimensional reduction is perfectly applicable
at the temperatures of the order of $g^2$, i.e., the critical ones (cf. the Introduction). The factor $\beta$ at the
action of the dimensionally-reduced theory, $S_{{\rm d.-r.}}=\beta\int d^2x{\cal L}[\chi,\psi|g_m,\zeta]$,
can be removed [and this action can be cast to the original form of eq.~(\ref{1}) with the substitution $d^3x\to d^2x$]
by the obvious rescaling:
$S_{{\rm d.-r.}}=\int d^2x{\cal L}\left[\chi^{\rm new},\psi^{\rm new}|\sqrt{T}g_m,\beta\zeta\right]$.
Here, $\chi^{\rm new}=\sqrt{\beta}\chi$, $\psi^{\rm new}=\sqrt{\beta}\psi$, and in what follows
we shall denote for brevity $\chi^{\rm new}$ and $\psi^{\rm new}$ simply as $\chi$ and $\psi$, respectively.
Averaging then over the field $\psi$ with the use of the cumulant expansion we arrive at the
following action:

$$
S_{{\rm d.-r.}}\simeq\int d^2x\left[\frac12(\nabla\chi)^2-2\xi\cos\left(g_m\sqrt{T}\chi\right)\right]-
$$

\begin{equation}
\label{2}
-2\xi^2\int d^2xd^2y\cos\left(g_m\sqrt{T}\chi({\bf x})\right){\cal K}({\bf x}-{\bf y})
\cos\left(g_m\sqrt{T}\chi({\bf y})\right).
\end{equation}
In this expression, we have disregarded all the cumulants higher
than the quadratic one, and the limits of applicability of
this so-called bilocal approximation will be discussed below.
In eq.~(\ref{2}), ${\cal K}({\bf x})\equiv{\rm e}^{g_m^2TD_{m_H}({\bf x})}-1$ with
$D_{m_H}({\bf x})\equiv K_0(m_H|{\bf x}|)/2\pi$ being the 2D Yukawa propagator
($K_0$ here is the modified Bessel function), and
$\xi\equiv\beta\zeta{\rm e}^{\frac{g_m^2T}{2}D_{m_H}(0)}$
denotes the monopole fugacity modified by the interaction of monopoles via the Higgs field.
Clearly, in the compact-QED limit (when $m_H$ formally tends to infinity) $D_{m_H}(0)$, being equal to
$\int\frac{d^2p}{(2\pi)^2}\frac{1}{p^2+m_H^2}$, vanishes already before doing the integration, and $\xi\to\zeta$,
as it should be. In the general case, when the mass of the Higgs field is moderate
and does not exceed $m_W$, which in the weak-coupling regime plays the r\^ole of the UV cutoff,
$\xi\propto\exp\left[-\frac{4\pi}{g^2}\left(m_W\epsilon+T\ln\left(\frac{{\rm e}^{\gamma}}{2}c\right)
\right)\right]$.
Here, we have introduced the notation $c\equiv m_H/m_W$, $c<1$, and $\gamma\simeq 0.577$ is the
Euler constant, so that $\frac{{\rm e}^{\gamma}}{2}\simeq 0.890<1$. We see that
the modified fugacity remains exponentially small, provided that

\begin{equation}
\label{3}
T<-\frac{m_W\epsilon}{\ln\left(\frac{{\rm e}^{\gamma}}{2}c\right)}.
\end{equation}
This constraint should be updated by another one, which would provide the convergence
of the cumulant expansion applied in the course of the average over $\psi$. In order to get this
new constraint, notice that the parameter of the cumulant expansion reads $\xi I$, where $I\equiv\int d^2x{\cal K}({\bf x})$.
The integral $I$ is evaluated in the Appendix~A and has the following form:

\begin{equation}
\label{I}
I\simeq\frac{2\pi}{m_H^2}\left[\frac12\left(c^2-1+\left(\frac{2}{{\rm e}^{\gamma}}
\right)^{\frac{8\pi T}{g^2}}\frac{1-c^{2-
\frac{8\pi T}{g^2}}}{1-\frac{4\pi T}{g^2}}\right)
+{\rm e}^{a/{\rm e}}-1+\frac{a}{{\rm e}}\right].
\end{equation}
(Note that at $T\to g^2/4\pi$, $\frac{1-c^{2-\frac{8\pi T}{g^2}}}{1-\frac{4\pi T}{g^2}}\to
-2\ln c$, i.e., $I$ remains finite.)
In the derivation of this expression, the parameter $a\equiv4\pi\sqrt{2\pi}T/g^2$ was assumed to be of the order of unity.
That is because the critical temperature of the deconfining phase transition, we are interested with, cannot
exceed the critical temperature of the monopole BKT phase transition, $g^2/2\pi$. In fact, above the point of the
BKT phase transition, the monopole ensemble passes to the molecular phase and loses its confining properties (in particular,
with respect to W's)~\footnote{In another words, $\xi$ vanishes together with $\zeta$ above the BKT critical
temperature. This is another reflection of the fact that confining strings disappear (i.e., become
infinitely thick) in that phase, since their thickness is proportional to $\xi^{-1/2}$.}.

Due to the exponential term in eq.~(\ref{I}),
the violation of the cumulant expansion may occur at high enough temperatures [that parallels the above-obtained
constraint~(\ref{3})].
The most essential, exponential, part of the parameter
of the cumulant expansion thus reads

$$\xi I\propto\exp\left[-\frac{4\pi}{g^2}\left(m_W\epsilon+T
\ln\left(\frac{{\rm e}^{\gamma}}{2}c\right)-T\frac{\sqrt{2\pi}}{{\rm e}}\right)\right].$$
Therefore, the cumulant expansion converges at the temperatures obeying the inequality

$$T<\frac{m_W\epsilon}{\frac{\sqrt{2\pi}}{{\rm e}}-\ln\left(\frac{{\rm e}^{\gamma}}{2}c\right)},$$
which updates the inequality~(\ref{3}). Since, as it has been just discussed in the preceding paragraph,
the temperatures we are working with do not exceed $g^2/2\pi$, it is enough to demand the following upper
bound on the parameter of the weak-coupling approximation, $g^2/m_W$:

$$\frac{g^2}{m_W}<\frac{2\pi\epsilon}{\frac{\sqrt{2\pi}}{{\rm e}}-
\ln\left(\frac{{\rm e}^{\gamma}}{2}c\right)}.$$
Note that although this inequality is satisfied automatically at $\frac{{\rm e}^{\gamma}}{2}c\sim 1$, since
it then takes the form $\frac{g^2}{m_W}<\sqrt{2\pi}{\rm e}\epsilon$, this is not so for the BPS limit,
$c\ll 1$. Indeed, in such a case,
we have $\frac{g^2}{m_W}\ln\left(\frac{2}{c{\rm e}^{\gamma}}\right)<2\pi\epsilon$,
that owing to the logarithm is however quite feasible.

\section{Critical temperatures of the deconfining phase transition}

We are now in the position to explore the influence of the Higgs-mediated interaction
of monopoles to the critical temperature of the deconfining phase transition. As it has already
been discussed in the Introduction, this phase transition occurs when the density of monopoles,
approximately equal to $2\xi$, becomes of the same order of magnitude as the density of W-bosons~\cite{W}.
The latter can be evaluated as follows (see e.g. ref.~\cite{f}):

$$\left.\rho_W=-\frac{\partial}{\partial\bar\mu}\left[6T\int\frac{d^2p}{(2\pi)^2}
\ln\left(1-{\rm e}^{\beta(\bar\mu-\varepsilon({\bf p}))}\right)\right]\right|_{\bar\mu=0}=
6\int\frac{d^2p}{(2\pi)^2}\frac{1}{{\rm e}^{\beta\varepsilon({\bf p})}-1}=$$

$$=\frac{3m_W^2}{\pi}\int\limits_{1}^{\infty}\frac{dzz}{{\rm e}^{m_W\beta z}-1}\simeq
\frac{3m_W^2}{\pi}\int\limits_{1}^{\infty}dzz{\rm e}^{-m_W\beta z}=
\frac{3m_WT}{\pi}\left(1+\frac{T}{m_W}\right){\rm e}^{-m_W\beta}.$$
Here, $\bar\mu$ stands for the chemical potential, $\varepsilon({\bf p})=\sqrt{{\bf p}^2+m_W^2}$, and
the factor "6" represents the total number of spin states of $W^{+}$- and
$W^{-}$-bosons. We have also denoted $z\equiv\varepsilon({\bf p})/m_W$ and
took into account that the temperatures
of our interest are much smaller than $m_W$ in the weak-coupling regime,
since they should not exceed $g^2/2\pi$.

Also, as it has been mentioned in the Introduction,
in the evaluation of the critical temperature(s), it is enough
to compare the exponents of $\xi$ and $\rho_W$,
since the preexponential factors yield only the subleading corrections.
Then, in the compact-QED limit, $\xi\to\zeta$ (cf. the preceding Section),
and $T_c=\frac{g^2}{4\pi\epsilon(\infty)}$~\cite{W}.
In the general case under discussion, $c<1$, we obtain
the two following distinct values of critical temperatures:

\begin{equation}
\label{Tc}
T_{1,2}=g^2\epsilon\frac{1\pm\sqrt{1-\frac{b}{\pi\epsilon^2}}}{2b}.
\end{equation}
Here, $b\equiv-\frac{g^2}{m_W}\ln\left(\frac{{\rm e}^{\gamma}}{2}c\right)$, $b>0$,
and the indices 1,2 refer to the smaller and the larger temperatures,
respectively.
The degenerate situation $T_1=T_2=g^2/2\pi\epsilon$ then corresponds to
$b=\pi\epsilon^2$,
and, since $\epsilon\ge 1$, $T_{1,2}\le g^2/2\pi$ in this case, as it should be.
In particular, in the
BPS limit, $\epsilon=1$, and the deconfining phase transition takes place together
with the monopole BKT one. Obviously, at any other $b<\pi\epsilon^2$, $T_1\ne T_2$, i.e., there exist
two separate phase transitions. (Note that the existence of the
upper bound for $b$ is quite
natural, since in the weak-coupling regime and aside from the
BPS limit, $b$ is definitely
bounded from above.) The existence of two phase transitions means that at $T=T_1$,
molecules of W-bosons start dissociating, while at $T=T_2$, this process is completed.
In another words, accounting for the interaction of monopoles via the Higgs field opens
a possibility for the existence of a new (metastable) phase at $T\in(T_1,T_2)$. This is the
phase, where both the gas of W-molecules and W-plasma are present.

An elementary analysis shows that for
$\pi(2\epsilon-1)<b\le\pi\epsilon^2$, $T_2<g^2/2\pi$ [and $T_2=g^2/2\pi$
at $b=\pi(2\epsilon-1)$]. At the values of $b$ lying in this interval,
the phase transition corresponding to the critical temperature $T_2$
thus may occur. In the BPS limit, $T_2$
can only be equal to $g^2/2\pi$, that corresponds to the above-discussed case when
both critical temperatures coincide with the one of the monopole BKT phase transition.
In the same way,
for any $b\le\pi\epsilon^2$, $T_1\le g^2/2\pi$, and, in particular,
$T_1=g^2/2\pi$ only in the BPS limit, when $\epsilon=1$. Therefore,
the phase transition corresponding to the temperature $T_1$ always takes place.
Also an elementary analysis shows that for any $\epsilon>0$
(and, in particular, for the realistic values $\epsilon\ge 1$)
and $b<\pi\epsilon^2$, $T_1>g^2/4\pi\epsilon$ (and consequently
$T_2>g^2/4\pi\epsilon$ as well). Since
$\epsilon<\epsilon(\infty)$, we conclude that both
phase transitions always occur at the temperatures which are larger than that
of the phase transition in the compact-QED limit.

Obviously, the RG analysis, performed in ref.~\cite{W} for the compact-QED limit remains valid, but
with the replacement $\zeta\to\xi$. In particular, the deconfining
phase transition corresponds again to the IR unstable fixed point,
where the exponent of the W-fugacity, $\mu\propto\rho_W$, is equal to
the exponent of $\xi$ [that yields the above-obtained
critical temperatures~(\ref{Tc})].
One can further see that the
initial condition $\mu_{\rm in}<\xi_{\rm in}$ takes place,
provided that the initial temperature, $T_{\rm in}$, is either smaller than $T_1$ or
lies between $T_2$ and $g^2/2\pi$. For these ranges of $T_{\rm in}$,
the temperature starts decreasing according to the RG equation $dt/d\lambda=
\pi^2\bar a^4\left(\mu^2-t^2\xi^2\right)$. In this equation, $t=4\pi T/g^2$, $\lambda$
is the evolution parameter, $\bar a$ is some
parameter of the dimensionality [length], and
for the comparison of $\mu$ and $\xi$ the preexponent $t^2$ is again immaterial. Then,
in the case $T_{\rm in}<T_1$, the situation is identical to the one
discussed in ref.~\cite{W}, namely $\mu$ becomes irrelevant
and decreases to zero. Indeed, from the evolution equation for $\mu$
there follows the equation for $d\mu/dt$, by virtue of which one can determine the sign
of this quantity. It reads

$$\frac{d\mu}{dt}=\frac{\mu\left(2-\frac1t\right)}{\pi^2\bar a^4\left(\mu^2-t^2\xi^2\right)}.$$
One can see from this equation that if the evolution starts at $T_{\rm in}\in(g^2/8\pi,T_1)$,
$\mu$ temporaly increases until the temperature is not equal to $g^2/8\pi$, but then nevertheless
starts vanishing together with the temperature.
However, by virtue of the same evolution equations we see that at $T_{\rm in}\in(T_2, g^2/2\pi)$,
the situation is now different.
Indeed, in that case, $\mu$ is not decreasing, but rather increasing with the decrease of the
temperature (since $d\mu/dt<0$ at $T>T_2$), until it reaches some
value $\mu_{*}\sim{\rm e}^{-m_W/T_2}$. Once the temperature becomes smaller than $T_2$,
the temperature starts increasing again, that together with the change of the sign of $d\mu/dt$  causes
the increase of $\mu$, and so on.
Thus, we see that $\mu_{*}$ is the stable local maximum of $\mu$ for such initial conditions.

\section{Conclusions}
In this letter, we have explored the consequences of accounting
for the Higgs field to the deconfining phase transition in the finite-temperature (2+1)D Georgi-Glashow
model. To this end, this field was not supposed to be infinitely
heavy, as it takes place in the compact-QED limit of the model.
Owing to that, the Higgs field starts propagating and, in particular,
causes the additional interaction of monopoles in the plasma.
This effect modifies the monopole fugacity, making it temperature-dependent, and
leads to the appearance of the novel terms in the action of the dual-photon field.
The cumulant expansion applied in the course of the average over the Higgs field
is checked to be convergent, provided that the weak-coupling approximation is implied in a certain sense.
Namely, the parameter of the weak-coupling approximation should be bounded from above
by a certain function of masses of the monopole, W-boson, and the Higgs field.

It has been demonstrated that although in the compact-QED limit there exists
only one critical temperature of the phase transition,
in general there exist two distinct critical temperatures.
We have discussed the dependence of these temperatures
on the parameters of the Georgi-Glashow model. In particular,
both critical temperatures turn out to be larger than the one of the phase
transition in the compact-QED limit. Besides that, it has been demonstrated that
the smaller of the two critical temperatures always does not exceed the critical
temperature of the monopole BKT phase transition. As far as the larger critical
temperature is concerned, there has been found the range of parameters
of the Georgi-Glashow model, where it also does not exceed the monopole one.
The situation when there exist two phase transitions implies that at the smaller
of the two critical temperatures, W-molecules
start dissociating, while at the larger one all of them are dissociated
completely. This means that in the region of temperatures between the
critical ones, the gas of W-molecules coexists with the W-plasma.

From the RG equations, it follows
that the presence of the second (larger) critical temperature
leads to the appearance of a novel stable value of the W-fugacity.
This value is reached if one starts the evolution in the region where the
temperature is larger than the above-mentioned critical one, and the density of W's is smaller than
the one of monopoles. The resulting stable value is nonvanishing (i.e., W's at that point are still
of some importance), that is the opposite to the standard situation, which
takes place if the evolution starts at the temperatures smaller than the first
critical one.

\section{Acknowledgments}
The author is greatful to Prof. A.~Di~Giacomo for useful discussions and cordial
hospitality and to Prof. I.I.~Kogan for a stimulating discussion and correspondence.
This work has been supported by INFN and partially by the INTAS grant Open Call 2000, Project No. 110. And last but not least,
the author is greatful to the whole staff of the Physics Department of the
University of Pisa for kind hospitality.

\section*{Appendix A. Evaluation of the integral $\int d^2x{\cal K}({\bf x})$.}

The desired integral can be written as follows:

$$
I=\frac{2\pi}{m_H^2}\int\limits_{c}^{\infty}dxx\left[\exp\left(\frac{8\pi T}{g^2}K_0(x)\right)-1\right]\simeq
$$

$$\simeq\frac{2\pi}{m_H^2}
\left\{\int\limits_{c}^{1}dxx\left[\exp\left(-\frac{8\pi T}{g^2}\ln\left(\frac{{\rm e}^{\gamma}}{2}x\right)
\right)-1\right]+
\int\limits_{1}^{\infty}dxx\left[\exp\left(a\frac{{\rm e}^{-x}}{\sqrt{x}}\right)-1\right]\right\}\equiv$$

$$
\equiv\frac{2\pi}{m_H^2}\left[\frac12\left(c^2-1+\left(\frac{2}{{\rm e}^{\gamma}}
\right)^{\frac{8\pi T}{g^2}}\frac{1-c^{2-\frac{8\pi T}{g^2}}}{1-\frac{4\pi T}{g^2}}\right)+
J\right],\eqno(A.1)
$$
where the notations $a$ and $c$ were introduced in the main text.
The integral $J$ here can further be evaluated as

$$J=\sum\limits_{n=1}^{\infty}\frac{a^n}{n!}\int\limits_{1}^{\infty}dx{\rm e}^{-nx}x^{1-\frac{n}{2}}=
\sum\limits_{n=1}^{\infty}\frac{a^n}{n!}n^{\frac{n}{2}-2}\Gamma\left(2-\frac{n}{2},n\right)
\simeq\sum\limits_{n=1}^{\infty}\frac{a^n}{nn!}{\rm e}^{-n}.\eqno(A.2)$$
Here, $\Gamma(a,x)=\int\limits_{x}^{\infty}dt{\rm e}^{-t}t^{a-1}$ is the incomplete Gamma-function,
and we have used its asymptotics $\Gamma(a,x)\simeq x^{a-1}{\rm e}^{-x}$ at $x\ge1$. One can further
evaluate the sum~(A.2) as follows:

$$
(A.2)=\int\limits_{0}^{\infty}dt\sum\limits_{n=1}^{\infty}a^n\frac{{\rm e}^{-(1+t)n}}{n!}\simeq\int\limits_{0}^{1}dt
\sum\limits_{n=1}^{\infty}a^n\frac{{\rm e}^{-n}}{n!}+\int\limits_{1}^{\infty}dt
\sum\limits_{n=1}^{\infty}a^n\frac{{\rm e}^{-tn}}{n!}={\rm e}^{a/{\rm e}}-1+\int\limits_{1}^{\infty}dt\left[
\exp\left(a{\rm e}^{-t}\right)-1\right]\simeq$$

$$\simeq{\rm e}^{a/{\rm e}}-1+a\int\limits_{1}^{\infty}dt{\rm e}^{-t}=
{\rm e}^{a/{\rm e}}-1+\frac{a}{{\rm e}}.$$
Inserting this expression into eq.~(A.1) we arrive at eq.~(\ref{I}) of the main text.


\newpage



\begin{thebibliography}{100}
\bibitem{1}A.M. Polyakov, Nucl. Phys. {\bf B 120} (1977) 429.
%%CITATION = NUPHA,B120,429;%%

\bibitem{2}
N.O. Agasian and K. Zarembo, Phys. Rev. {\bf D 57} (1998) 2475.
%%CITATION = PHRVA,D57,2475;%%

\bibitem{W}
G. Dunne, I.I. Kogan, A. Kovner, and B. Tekin, JHEP {\bf 01} (2001) 032.
%%CITATION = JHEPA,0101,032;%%

\bibitem{W1}
I.I. Kogan, A. Kovner, and B. Tekin,
JHEP {\bf 03} (2001) 021;
Phys. Rev. {\bf D 63} (2001) 116007;
JHEP {\bf 05} (2001) 062;
I.I. Kogan, A. Kovner, and M. Schvellinger,
JHEP {\bf 07} (2001) 019.
%%CITATION = JHEPA,0103,021;%%
%%CITATION = JHEPA,0105,062;%%
%%CITATION = JHEPA,0107,019;%%
%%CITATION = PHRVA,D63,116007;%%

\bibitem{q}
N. Agasian and D. Antonov, Phys. Lett. {\bf B 530} (2002) 153.
%%CITATION = HEP-TH 0109189;%%

\bibitem{nd}
N. Agasian and D. Antonov, JHEP {\bf 06} (2001) 058
[for short reviews see: D. Antonov, talks given at the 6th Workshop on nonperturbative QCD
(Paris, France, 5-9th June, 2001) and at the 10th Lomonosov conference on elementary particle
physics (Moscow, Russia, 23-29th August, 2001):
preprints {\tt hep-th/0109071} and
{\tt hep-th/0111223}].
%%CITATION = JHEPA,0106,058;%%
%%CITATION = HEP-TH 0109071;%%
%%CITATION = HEP-TH 0111223;%%


\bibitem{BKT}
V.L. Berezinsky, Sov. Phys.- JETP {\bf 32} (1971) 493;
J.M. Kosterlitz and D.J. Thouless, J. Phys. {\bf C 6}
(1973) 1181; J.M. Kosterlitz, J. Phys. {\bf C 7} (1974) 1046.
%%CITATION = JTPHE,32,493;%%
%%CITATION = JPCBA,6,1181;%%
%%CITATION = JPCBA,7,1046;%%

\bibitem{dietz}
K. Dietz and Th. Filk, Nucl. Phys. {\bf B 164} (1980) 536.
%%CITATION = NUPHA,B164,536;%%

\bibitem{bps}
M.K. Prasad and C.M. Sommerfield, Phys. Rev. Lett. {\bf 35} (1975) 760;
E.B. Bogomolny, Sov. J. Nucl. Phys. {\bf 24} (1976) 449.
%%CITATION = PRLTA,35,760;%%
%%CITATION = YAFIA,24,861;%%


\bibitem{kirk}
T.W. Kirkman and C.K. Zachos, Phys. Rev. {\bf D 24} (1981) 999.
%%CITATION = PHRVA,D24,999;%%

\bibitem{ks}
V.G. Kiselev and K.G. Selivanov, Phys. Lett. {\bf B 213} (1988) 165.
%%CITATION = PHLTA,B213,165;%%

\bibitem{f}
R.P. Feynman, {\it Statistical mechanics. A set of lectures} (Addison-Wesley Publishing Company, Inc., The Advanced Book
Program, Reading, Massachusetts, 1972).


\end{thebibliography}
\end{document}